\documentclass[preprint,showpacs,preprintnumbers,amsmath,amssymb]{revtex4}
\usepackage{graphicx}
\usepackage{dcolumn}
\usepackage{bm}

\begin{document}
\preprint{APS}

\title{Scale-free nonlinear conservative cascades \\ and their stationary spectra}

\author{Dmitri O. Pushkin}
\email{pouchkin@uiuc.edu}
\affiliation{Department of Theoretical
and Applied Mechanics, University of Illinois at Urbana-Champaign,
Urbana, IL 61801, USA}

\author{Hassan Aref}
\email{haref@vt.edu}

\affiliation{ Virginia Polytechnic Institute \& State University,
Blacksburg, VA 24061-0217, USA }

\date{March 26, 2004}

\begin{abstract}
We show that a variety of complex processes can be viewed from the
unified standpoint of scale-free nonlinear conservative cascades.
Examples include certain turbulence models, percolation, cluster
coagulation (aggregation) and fragmentation, `coarse-grained'
forest fire model of self-organized criticality, and scale-free
network growth. We classify such cascades by the values of three
indices, and show how power-law steady spectra may arise. The
power-law exponent is proven to depend only on the values of the
three indices by a simple algebraic formula.

\end{abstract}

\maketitle

The cascade idea is central to our understanding of an amazingly
broad scope of natural phenomena. One of the earliest mathematical
cascade models, the Galton-Watson process \cite{Wa1874}, was
suggested in 1874 as a solution to the "problem of the extinction
of families." This kind of branching cascades was later used for
theoretical treatment of chemical and nuclear chain reactions
\cite{Se35}, cosmic radiation \cite{Bha37}, and statistics of
polymers \cite{Go61} among other physical and biological problems.

Cluster fragmentation is another common type of cascades
\cite{Ko41}. Typical examples of fragmentation are rock fracture,
breakup of liquid droplets \cite{Shi61}, and polymer degradation
\cite{Zi85}. Fragmentation preserves the total mass of clusters.
Both branching processes and fragmentation are linear cascades, as
the agents are assumed to not interfere with each other.

Studies of turbulence have provided a wealth of examples of
nonlinear cascades. Among them are the Richardson cascade of
energy in the inertial range that underlies the now classical
approach of Kolmogorov \cite{Ko41b}, the Kraichnan reverse
enstrophy cascade for two-dimensional turbulence \cite{Kra67}, and
various random cascade models suggested as theoretical
explanations of the unsolved problem of turbulence intermittency
\cite{Ma74}.

More recently the cascade idea has been frequently evoked in
connection with a variety of complex processes. A self-organized
branching cascade process was put forth as a mean-field theory for
avalanches \cite{Za95}. A cascade model was suggested as a
`coarse-grained' approximation to the forest fire model,
displaying self-organized criticality \cite{Tu99}.  The classic
result for the asteroidal size distribution \cite{Do69} has been
re-considered from the perspective of collisional fragmentation
cascades \cite{Wi94}. Mesoscopic rainfall \cite{Gu93}, cascades of
reconnecting magnetic loops in solar flares \cite{Hu03}, and
cascades of data networks \cite{Fe98} are further examples from
this vast list.

We remark, however, that other important complex systems and
processes -- e.g. multiplicative random processes, percolation
\cite{Sta85}, cluster coagulation \cite{Pu04}, Scheidegger model
of river networks, directed Abelian sandpile model \cite{Dha99},
and scale-free random networks \cite{Ba99} -- can be
advantageously considered as nonlinear cascades.

Nonlinear cascades are characterized by strong interaction between
the agents. Unlike linear cascades, they are are poorly
understood. Their studies have usually relied on system-specific
features and assumptions and, as a result, the general picture has
remained unknown.

The purpose of the present paper is to study a general class of
scale-free nonlinear conservative cascades, and to show that
seemingly different systems reveal common features when considered
from this unified viewpoint. In particular, we derive a nonlinear
cascade equation, and demonstrate how it can give rise to
stationary power-law spectra. We show that the power-law exponent
depends only on three indices characterizing a particular cascade:
the `conservation law' index $m$, the {\em scale homogeneity
index} $\alpha$, and the {\em nonlinearity index} $h$. This result
is illustrated by Table 1.
\begin{table*}
\caption{ Some scale-free conservative cascades with stationary
power-law spectra, $n(s) \propto s^{-\tau}$, discovered in various
contexts. For all of them $\tau = 1 + (m + \alpha)/h$.}
\begin{ruledtabular}
\begin{tabular}{cccccc} Cascade type & $h$ & $m$ & $\alpha$ & $\tau$ & Notes
\\
\hline
Turbulent energy    & $3/2$     & $0$   & $1$   & $5/3$     &
Kolmogorov (1941) \cite{Ko41b}\\     Enstrophy   & $3/2$     & $2$
& $1$ & $3$ & Kraichnan (1967) \cite{Kra67}\\    Passive tracer &
$1$ & $0$ & $0$ & $1$       & Batchelor (1959) \cite{Ba59} \\

Percolation & $2$   & $1$   & $2$   & $5/2$ & \\

Diffusion limited cluster aggregation & $2$ &   $1$  &   $0$  &
3/2 &    \\

Cluster coagulation in a shear flow &   $2$ &   $1$  &   $1$  &
$2$ &   Hunt(1982) \cite{Hu82}\\

Scale-free cluster coagulation  & $2$   & $1$ & $\alpha$ &
$(3+\alpha)/2$ & Pushkin and Aref (2002) \cite{Pu02} \\

Cascade model of forest fires &  $2$  & $1$  & $1$  & $2$ &
Turcotte (1999) \cite{Tu99}\\

Directed Abelian sandpile model & $3$   & $1$ & $0$ & $4/3$   &
Dhar and Ramaswami (1989) \cite{Dha99}\\

1D-diffusion limited cluster aggregation    & $3$   & $1$ & $0$ &
$4/3$   & Takayasu (1989) \cite{Ta89}\\

Asteroid collisional fragmentation   & $2$   & $1$   & $2/3$ &
$11/6$ & Dohnanyi (1969) \cite{Do69}\\

Scale-free collisional fragmentation   &   $2$     &   $1$     &
$\alpha$ & $(3+\alpha)/2$  &   Tanaka et. al. (1996) \cite{Wi94}
\\

`Random scission' fragmentation & $1$   & $1$ & $1$   & $3$ &
Ben-Naim and Krapivsky (2000) \cite{Be00} \\

Scale-free random networks & $1$   & $1$   & $1$   & $3$   &
Barabasi and Albert (1999) \cite{Ba99}

\end{tabular}
\end{ruledtabular}
\end{table*}

Let us set the stage. We will think of cascades as statistical
collective systems of interacting agents of different type.
Examples of agents are turbulent eddies, coalescing rain droplets,
avalanches, connected network clusters and polymer chains, genetic
family trees, etc. As a result of an interaction of two or more
agents, `new' agents form, whereas `old' agents annihilate. For
instance, the result of hydrodynamic interactions between
turbulent eddies leading to exchange of their energies, is viewed
as annihilation of `old' eddies and creation of `new' eddies
having altered energies in their place. Any such interaction can
be written formally as
\begin{eqnarray}
\mathrm{A}_{s_1} + \mathrm{A}_{s_2}+... \to \mathrm{A}_{s'_1} +
\mathrm{A}_{s'_2}+...\; ; \;\; \mbox{ or  } \;\; \sum_{s_i}
\nu_{s_i} \mathrm{A}_{s_i} = 0. \label{eq:reaction}
\end{eqnarray}
Here $s$ labels various agent types, and can be discrete or
continuous. In the latter case summation has to be substituted by
integration. Here we will assume $s$ to be a real number, and
refer to it as `size'. $ \nu_{s_i} $'s are taken positive for
`new' agents and negative for `old' ones. Although this notation
is routinely used for chemical reactions, it is quite unusual in
other fields, e.g. turbulence.

We assume that agents of the same type are indistinguishable in
statistical sense, and that a broad distribution of agent types
emerges as a result of interactions.

A cascade state is specified by the agent distribution, also
called spectrum, $n(t,s)$. Common examples of agent distributions
are the turbulent energy spectrum $E(t,k)$ in homogeneous
isotropic turbulence, where the scalar wavenumber $k$ plays the
role of $s$; the avalanche size distribution $n(t,s)$; and the
distribution of masses of clusters formed due to cluster
coagulation or/and fragmentation.

The primary goal of a cascade theory is to predict the spectrum
$n$. We derive a novel formal nonlinear cascade equation, which
describes temporal evolution of $n$, from the master equation
\cite{Ka81}:
\begin{equation}
\dot{P}_t[N(s)] = \sum_{N_1(s)} w[N(s),N_1(s)] P_t[N_1(s)] -
w[N_1(s),N(s)] P_t[N(s)]. \label{eq:master}
\end{equation}
Here $P_t[N(s)]$ is the time-dependent probability of the
microstate $N$, square brackets emphasize functional dependence;
$w[N,N_1]$ is the probability of transition from the microstate
$N_1$ to the microstate $N$ per unit time; summation is carried
over all microstates compatible with the interactions
(\ref{eq:reaction}). Then the macroscopic spectrum $n(t,s)$ can be
obtained as an average over all microstates:
\begin{equation}
n(t,s) = \left<  N(s) \right>  \equiv \sum_N N(s) P_t[N].
\label{eq:n}
\end{equation}
After multiplying (\ref{eq:master}) by $N(s)$, summing it over all
microstates $N$, regrouping the terms, and introducing $\Delta
N(s) \equiv N(s)-N_1(s)$, we obtain
\begin{eqnarray}
\dot{n}(t,s) = \sum_{\Delta N} \Delta N(s) \sum_{N} w[N + \Delta
N, N] P_t[N]. \nonumber
\end{eqnarray}
When an interaction is of the type (\ref{eq:reaction}), $\Delta
N(s)$ has the form:
\begin{equation}
\Delta N(s) = \nu(s, \{ s_i \}) \equiv \sum_{ s_i } \nu_{s_i}
\delta(s-s_i). \label{eq:deltaN}
\end{equation}
Clearly, $\nu(s, \{ s_i \})$ is the number of agents of size $s$
created due to an `elementary' interaction of agents of sizes $\{
s_i \} $. Now we can define the interaction probability per unit
time:
\begin{equation}
W ( \{ s_i \}, [N(s)] ) = w[N(s) + \nu(s, \{ s_i \}), N(s) ],
\label{eq:new_w}
\end{equation}
and its average:
\begin{equation}
\mathcal{T}(t,\{ s_i \}) \equiv \left< \, W(\{ s_i \}, [N]) \,
\right>. \label{eq:interact}
\end{equation}
The physical sense of $\mathcal{T}$ is interaction strength
between the agents of sizes $\{s_i\}$. Thus, we arrive at the {\em
nonlinear cascade equation}:
\begin{equation}
\dot n(t,s) = \left< \left< \;  \nu(s, \{ s_i \} ) \; \right>
\right>, \label{eq:cascade}
\end{equation}
where the following notation has been introduced: for any function
$f(\{ s_i \})$,
\begin{eqnarray}
\left< \left< \;  f(\{ s_i \}) \; \right> \right> (t) \equiv
\sum_{ \{ s_i \} } f(\{ s_i \}) \mathcal{T}(t,\{ s_i \}).
\label{eq:brackets}
\end{eqnarray}
The summation in (\ref{eq:brackets}) is carried over all sets $\{
s_i \}$. At this point equation (\ref{eq:cascade}) is not closed,
as the interaction strength $\mathcal{T}$ depends on the
microdistribution $N$. The necessary closure can be provided for
particular types of interactions by mean field-type approximations
or renormalization procedures among other methods \cite{Go92}.
However, we will show that for conservative scale-free cascades
important conclusions about asymptotic spectra can be found
by-passing the closure problem.

Let us illustrate the above ideas with two examples. For cluster
coagulation \cite{Pu04}:
\begin{eqnarray}
\mathrm{A}_{s_i} + \mathrm{A}_{s_j} &\to& \mathrm{A}_{s_{i + j}},
\nonumber\\ \nu(s;s_i,s_j,s_i+s_j) &=& \delta_{s,s_i+s_j} -
\delta_{s, s_i} - \delta_{s, s_j,} \nonumber\\ W( t; s_i, s_j,
s_{i + j} ; [N]) &=& K(s_i,s_j)N_{s_i} N_{s_j}, \; \mbox{for any
$i,j$}. \nonumber
\end{eqnarray}
Here $K(i,j)$ is the probability that two cluster of sizes $s_i$
and $s_j$ aggregate per unit time. $K$ is commonly called
coagulation kernel. In the mean-field approximation:
\begin{equation}
\mathcal{T} = K(s_i,s_j)n_{s_i} n_{s_j}, \nonumber
\end{equation}
and (\ref{eq:cascade}) becomes the celebrated Smoluchowski
coagulation equation:
\begin{equation}
\dot n_{s}(t) = \frac12 \sum_{s_i=1}^{s-1} K(s_i,s-s_i) n_{s_i}(t)
n_{s-s_i}(t) - n_{s}(t) \sum_{s_i=1}^{\infty} K(s,s_i) n_{s_i}(t).
\label{eq:SCE}
\end{equation}

As another example consider two-dimensional isotropic turbulence.
In the inviscid limit the energy balance equation for a turbulent
flow in a cyclic box of side $D$ reads \cite{Kra67}:
\begin{subequations}
\begin{eqnarray}
\partial_t E(k) &=& T(k), \;
T(k) = \frac12 \int_0^{\infty}  dp \, dq \, T(k,p,q), \nonumber
\end{eqnarray}
where
\begin{eqnarray}
 T(k,p,q) &=& 2 \pi k \mbox{Im} \{ (D/2 \pi)^4 (2 \pi/| \sin(\mathbf p,
\mathbf q)|) \nonumber\\ &\times& (k_m \delta_{ij}+ k_j
\delta_{im}) \left< u_i^{*}(\mathbf k) u_j(\mathbf p) u_m(\mathbf
q) \right> \}, \nonumber\\ && ( \mathbf k = \mathbf p + \mathbf q,
\, k = |\mathbf k|, \, p = |\mathbf p|, \, q = |\mathbf q|).
\nonumber
\end{eqnarray}
\end{subequations}
This problem can be easily written in the form (\ref{eq:cascade}):
\begin{eqnarray}
\mathrm{A}_k + \nu_p \mathrm{A}_p &+& \nu_q \mathrm{A}_q = 0,
\nonumber\\ \nu(s;k,p,q) = \delta(s-k) &+& \nu_p \delta(s-p) +
\nu_q \delta(s-q), \nonumber\\ \nu_p = T(p,k,q)/T(k,p,q), && \,
\nu_q = T(q,p,k)/T(k,p,q), \nonumber\\
\mathcal{T}(k,p,q)&=&T(k,p,q). \nonumber
\end{eqnarray}

The condition that a cascade is scale-free has several
consequences: first, the reaction type defined by
(\ref{eq:reaction}) is independent of the physical scale of agent
sizes. More exactly, any two sets of agent sizes, $\{ s_i \}$ and
$\{ s_i' \}$, such that $s_i' = \lambda s_i$ for some positive
$\lambda$, allow reactions of the same type, i.e. $\nu_{s_i} =
\nu_{s_i'}$.

Next, interactions should have no characteristic size. In
mathematical terms: if $s_i'= \lambda s_i, \; N' = \lambda^{-1}
N$, then
\begin{equation}
\left< \, W( \{ s_i' \}, [N']) \, \right> = \lambda^{\alpha}
\left< \, W( \{ s_i \}, [N] ) \, \right>. \label{eq:alpha}
\end{equation}
For instance, for homogeneous isotropic turbulence $\alpha=1$ and
this value can be traced to the gradient operator in the
Navier-Stokes equations. For coagulation, $\alpha$ is the
homogeneity degree of the coagulation kernel.

Finally, the interactions must be scalable in agent density. Thus,
if $s_i'=s_i, \; N' = \lambda N$,
\begin{equation}
\left< \, W( \{ s_i' \}, [N']) \, \right> = \lambda^{h} \left< \,
W( \{ s_i \}, [N] ) \, \right>. \label{eq:h}
\end{equation}
For example, for homogeneous isotropic turbulence $h=3/2$, because
the interactions depend on the third order velocity correlation
function, and velocity scales as square root of energy. For
coagulation, $h=2$ due to binary collisions.

Let the conservation law read:
\begin{equation}
\sum_{s_j} \nu(s_j, \{ s_i \}) s_j^m = 0, \; \mbox{for any set $\{
s_i \}$.} \label{eq:conserv}
\end{equation}
For instance, the energy conservation law for turbulence yields
$m=0$, while enstrophy conservation for 2D turbulence results in
$m=2$. Mass conservation in coagulation and fragmentation cascades
yields $m=1$.

The nonlinear cascade equation can be used for treating open
systems, e.g. a constantly stirred turbulent flow, a coagulating
system with a source of the smallest clusters, or a driven
self-organizing system. Let us assume (without limitation of
generality) a direct cascade, i.e. $s>s_0$, where $s_0$ is the
smallest size in the system. Then, we supplement equation
(\ref{eq:cascade}) with the `boundary condition':
\begin{equation}
n(t,s_0)=n_0 \; \mbox{for all $t$},  \label{eq:BC}
\end{equation}
and look for stationary solutions. We are particularly interested
in the limit $s \gg s_0$. For scale-free conservative cascades
this problem can be easily solved in a general way. Because the
conserved quantity flux, $E$, is well-defined in the limit $s/s_0
\to \infty$, one can expect that solutions of (\ref{eq:cascade}),
(\ref{eq:BC}) depend on $E$ and `forget' the microscale $s_0$.
These expectations are corroborated below.

It follows from (\ref{eq:cascade}) and the definition of $E$ that:
\begin{equation}
E =  \sum_{ \{ s_i \} } \left< \, W( \{ s_i \}, [N]) \, \right>
\sum_{s_i} \nu_{s_i} s_i^m \theta (S-s_i), \; \mbox{for any
$s_0<S<\infty$} . \label{eq:E}
\end{equation}
Here $\theta(x)$ is the step function. A lot of instances of
scale-free conservative cascades are known to evolve steady
power-law spectra. Now we can understand how they arise. Indeed,
it follows from the definition of spectrum (\ref{eq:n}), the
properties (\ref{eq:alpha}) and (\ref{eq:h}), and the
arbitrariness of $S$ in (\ref{eq:E}), that
\begin{eqnarray}
n(s,s_0,E) &=& E^{1/h} \lambda^{\tau} n(\lambda s,\lambda s_0,1),
\; \mbox{for any positive $\lambda$}, \nonumber\\ \tau &=& 1 +
\frac{m+\alpha}{h}. \label{eq:tau}
\end{eqnarray}
Hence $n$ has the form:
\begin{eqnarray}
n(s,s_0,E) = E^{1/h} s^{-\tau} f(s_0/s). \label{eq:scale_f}
\end{eqnarray}
For wide classes of interactions the right hand side of
(\ref{eq:E}) has a finite limit as $s_0/S \to 0$. These classes
depend on asymptotic properties of interactions. They have been
determined for cluster coagulation \cite{Pu02} and weak wave
turbulence (`locality conditions') \cite{Za}. In this limit the
scale $s_0$ drops out of the equations, and the function $f$ has a
finite limit:
\begin{eqnarray}
\lim_{x \to 0} f(x) = C, \;\; \mbox{and, thus,} \;\; n(s) \approx
C E^{1/h} s^{- \tau}. \label{eq:spectrum}
\end{eqnarray}
In turbulence the constant $C$ was first introduced by Kolmogorov
\cite{Ko41}, and is often denoted as $C_2$. It is clearly
interaction-dependent and, therefore, non-universal. For cluster
coagulation an expression for $C$ was found in \cite{Pu02}.

At this point a remark is appropriate: the choice of size
variable, which one faces looking for a model for a cascade
process, is quite arbitrary. For example, one could characterize
cluster coagulation by the cluster diameter distribution, rather
than the cluster mass distribution. Let the new size variable
$\hat s = s^q.$ Then $\hat n(\hat s)d \hat s= n ( s) d s$, and
$(\tau-1) \to q (\tau - 1)$. Clearly, (\ref{eq:tau}) must obey
this transformation rule. It does so, indeed, as under this
transform $ \{ m ,\ \alpha ,\ h \} \to \{ q m ,\ q \alpha ,\ h \}
$.

The self-similar distribution (\ref{eq:spectrum}) is our central
result. Table 1 demonstrates that for many complex processes,
which can be viewed as scale-free conservative cascades, the
exponents of power-law spectra can be obtained in nearly automatic
fashion. More of examples of self-similar spectra can doubtlessly
be found scattered throughout the scientific literature in various
fields. Because values of the indices for a particular cascade are
often apparent due to their clear physical meaning, our result has
considerable heuristic powers.

The semblance between phenomena in the fields of turbulence,
cluster coagulation (aggregation) and fragmentation,
self-organized criticality, critical phenomena, and, more
recently, econophysics -- has long been perceived, e.g.
\cite{Ey94}. In certain instances it has led to new quantitative
results: e.g. the self-similar spectra due to cluster coagulation
have been derived for three coagulation mechanisms -- Brownian,
laminar shear, and sedimentation in gravity -- using reasoning
patterned on the Kolmogorov 1941 theory of turbulence \cite{Hu82}.
Our approach justifies and generalizes such results.

It is notable that most of well-studied cascade processes --
branching processes, fragmentation, (strange/anomalous)
diffusions, L\'evy flights, multiplicative processes, and cascade
models of turbulence intermittency -- are linear cascades with
$h=1$. Although just a particular class of scale-free cascades,
their investigation has led to rich developments in our
understanding and modeling of complex phenomena. We expect,
therefore, that investigation of nonlinear cascades is essential
to further progress in this field.

\end{document}